# Correlated electrons of the flat band in charge density wave state of $4H_b$-TaSe$_x$S$_{2-x}$

Yanyan Geng[1,2,+], Jianfeng Guo[1,2,+], Fanyu Meng[1,2,+], Manyu Wang[1,2], Shuo Mi[1,2], Li Huang[3], Rui Xu[1,2], Fei Pang[1,2], Kai Liu[1,2], Shancai Wang[1,2], Hong-Jun Gao[3], Weichang Zhou[4], Wei Ji[1,2,*], Hechang Lei[1,2,*], and Zhihai Cheng[1,2,*]

[1]Key Laboratory of Quantum State Construction and Manipulation (Ministry of Education), Renmin University of China, Beijing, 100872, China

[2]Beijing Key Laboratory of Optoelectronic Functional Materials & Micro-nano Devices, Department of Physics, Renmin University of China, Beijing 100872, China

[3]Beijing National Laboratory for Condensed Matter Physics, Institute of Physics, Chinese Academy of Sciences, Beijing 100190, China

[4]Key Laboratory of Low-dimensional Quantum Structures and Quantum Control of Ministry of Education, School of Physics and Electronics, Institute of Interdisciplinary Studies, Hunan Normal University, Changsha 410081, China

**Abstract:** Many intriguing quantum states of matter, such as unconventional superconductivity, magnetic phases and fractional quantum Hall physics, emergent from the spatially-correlated localized electrons in the flat band of solid materials. By using scanning tunneling microscopy and spectroscopy (STM/STS), we report the real-space investigation of correlated electrons in the flat band of superlattice $4H_b$-TaSe$_x$S$_{2-x}$. In contrast with the pristine $4H_b$-TaS$_2$, the selenium (Se) substitutions significantly affect the interfacial transfer of correlated electrons between the CDW states of 1T- and 1H-TaS$_2$ layers, and contribute a real-space fractional electron-filling configurations with the distributed electron-filled and -void SoD clusters of 1T-layer. The site-specific STS spectra directly reveal their respective prominent spectra weight above $E_F$ and symmetric Mott-like spectra. In addition, the spatial distributions of these electron-filled SoDs in the 1T-layer of $4H_b$-TaSe$_{0.7}$S$_{1.3}$ demonstrate different local short-range patterning, clearly indicating the complex neighboring interactions among the localized electrons in the flat band of 1T-layer. Our results not only provide an in-depth insight of correlated electrons in the flat CDW band, and provide a simple platform to manipulate the electron-correlation-related quantum states.

[+]These authors contributed equally: Yanyan Geng, Jianfeng Guo, Fanyu Meng,

*Corresponding author. E-mail address: zhihaicheng@ruc.edu.cn, hlei@ruc.edu.cn wji@ruc.edu.cn

**Introduction**

Flat bands, characterized by their strong correlation arising from the high density of states and their dispersionless nature, have gained considerable attention from both condensed matter physics and materials science. As a special electron band structure, the flat band has a large number of energy-degenerate electrons, which can still be confined in real-space in lattices supporting dispersionless electronic excitation in momentum space. Due to electron-electron interactions above the quenched kinetic energy, materials with electronic flat bands provide an attractive ground to explore various exotic phenomena, including unconventional superconductivity [1-3], magnetic phases [4,5], fractional quantum Hall physics [6], excitonic insulating behavior [7,8], and charge density wave (CDW) states [9-11]. Recent advances in kagome lattice [12,13], twisted bilayer graphene and twisted transition metal dichalcogenides (TMDs) in moiré superlattice systems [9,14], in some TMDs with CDW [15,16], prompted researchers to explore the realization of flat bands in a broader range of material systems.

$TaS_2$ is one of the most studied CDW material with rich crystal structures and physical properties [17,18]. The 1$T$-$TaS_2$ exhibits a strong-correlation Mott-insulating ground state in the commensurate CDW (CCDW) state [19]. In the CCDW state, every 13 Ta atoms shrink into a cluster named as the Star of David (SoD). Then, the SoD clusters arrange into an ordered triangular $\sqrt{13} \times \sqrt{13}$ superlattice, as shown in Fig. 1(a). Each SoD contains 13 Ta 5d electrons, the 12 electrons of the outer Ta atoms pair and form six occupied insulating bands, and leave one unpaired electron (Mott) of the central Ta atom in a half-filled metallic band at $E_F$. This half-filled band further splits into upper and lower Hubbard bands (UHB and LHB) due to the large onsite Coulomb interaction U and hence forms a Mott insulator [Fig. S1]. However, recent studies demonstrate that the interlayer CDW dimerization results in the formation of a band insulator rather than a correlated one [20-22]. While, 2$H$-$TaS_2$ exhibits metallic behavior with a coexisting commensurate 3 × 3 CDW at ~78 K [Fig. 1(b)] and superconductivity at ~0.8 K [23]. The typical STM images of the $\sqrt{13} \times \sqrt{13}$ CDW (1T-layer) and 3 × 3 CDW (2H-layer) are shown in Figs. 1(c) and 1(d).

Using the van der Waals nature, $TaS_2$ can be spontaneously assembled into a superlattice with alternating insulating T phase and metallic H phase, known as the 4$H_b$ phase [Fig. 1(e)] and 6R phase [24,25]. $4H_b$-TaS$_2$ has evoked great interest owing to the ability to induce new electronic ground states not present in T phase or H phase, which profoundly alter in both its superconducting and Mott ground states. Recent studies have revealed the existence of the chiral superconducting phase [26], two-component nematic superconductivity [27], and topological nodal superconductivity of the 1H-layer [28]. Scanning tunnelling spectroscopy (STS) measurements reported the narrow electronic bands above the $E_F$ of $4H_b$-TaS$_2$ [29], and the Kondo effect on 1T/1H bilayers by MBE grown [30] or some specific CDW sites of $4H_b$-TaS$_2$ [31]. In addition to their intrinsic states, many exotic states can be induced by electric field [31], temperature [32], substrate [33], and so on. Precise isovalent Se substitution is also an effective method, which does not introduce extra electron or hole into the system. The study of temperature-dependent resistivity showed that the superconducting transition temperature is highest (~ 4.1 K) and the 3× 3 CDW of 1H-layer at ~22 K at the optimal Se substitution content of 0.7 [34]. Significantly, Se substitution can alter the interlayer distance [32], which can effectively modify the interlayer charge transfer of electrons [35]. Most studies have focused on the effect of Se substitution on superconductivity and transport properties [32, 34], however, the investigation on the effect of Se substitution on CDW, interfacial electron transfer and real-space electron-filling configurations are rare and still challenging.

In this paper, the real-space investigation of correlated electrons in the flat band of superlattice $4H_b$-TaSe$_x$S$_{2-x}$ are reported by STM/STS. Compared with the pristine $4H_b$-TaS$_2$, the effects of Se substitution on the work function (surface potential) difference, CDW order, interfacial transfer of correlated electrons of 1T- and 1H-TaS$_2$ layers, and spatial fractional electron-filling configurations with the distributed electron-filled and -void SoD clusters of 1T-layer are given in detail. The site-specific STS spectra further reveal their symmetric Mott-like spectra and respective prominent spectra weight above E$_F$ of electron-filled and -void SoD clusters of 1T-layer. Moreover, we demonstrate different local short-range patterning of these electron-filled SoDs in the 1T-layer of $4H_b$-TaSe$_{0.7}$S$_{1.3}$, which can be attributed to the complex neighboring interactions among the localized electrons in the flat band of 1T-layer.

# Results and discussion

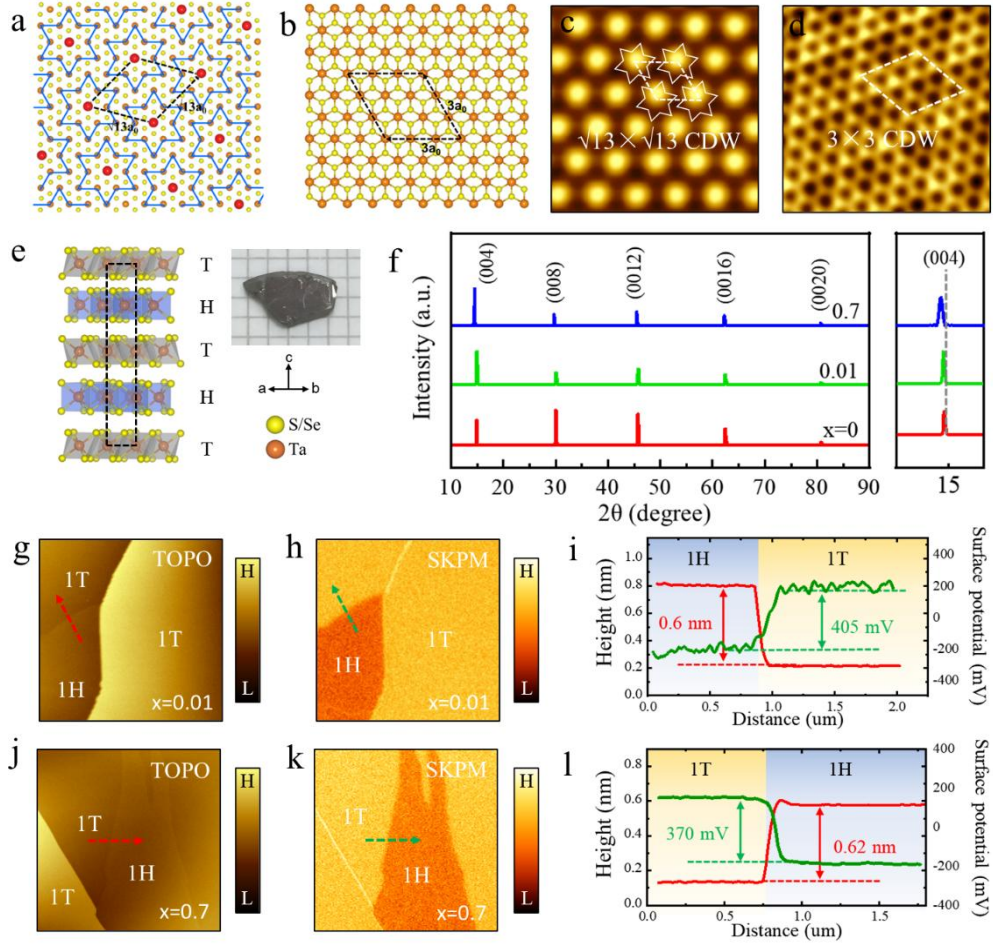

**Figure 1. Crystal structure and SKPM characterization of superlattice 4$H_b$-TaSe$_x$S$_{2-x}$.** (a,b) Schematic illustrations of the 1T-layer (a) and 2H-layer (b). The supercells of the √13×√13 CDW (1T-layer) and 3×3 CDW (2H-layer) are marked by the black dashed lines in (a) and (b), respectively. The SoD clusters of the √13×√13 CDW state in 1T-layer are outlined by the blue stars with the central localized electrons of CDW flat band. (c, d) The typical STM images of 1$T$-TaS$_2$ (c) and 2$H$-TaS$_2$ (d). The √13×√13 CDW (1T-layer) and 3×3 CDW (2H-layer) are clearly visible. (e) Crystal structure and typical optical image of the 4$H_b$-TaSe$_x$S$_{2-x}$ single crystal. The Ta and S/Se atoms are shown in orange and yellow, respectively. (f) XRD patterns of 4$H_b$-TaS$_2$, 4$H_b$-TaSe$_{0.01}$S$_{1.99}$, and 4$H_b$-TaSe$_{0.7}$S$_{1.3}$. (g,h) AFM topography (g) and SKPM surface potential (h) images of 4$H_b$-TaSe$_{0.01}$S$_{1.99}$. (i) Line-profile along the dashed line in (g) and (h). The step height and surface potential difference between 1T-layer and 1H-layer are ~0.6 nm and ~400 mV, respectively. (j, k) AFM topography (j) and SKPM surface potential (k) images of 4$H_b$-TaSe$_{0.7}$S$_{1.3}$. (l) Line-profile along the dashed line in (j) and (k). The step height and surface potential difference between 1T-layer and 1H-layer are ~0.6 nm and ~370 mV, respectively. Imaging parameters: (a) 6 × 6 nm$^2$; (b) 3 × 3 nm$^2$; (f,g,i,j) 5 × 5 um$^2$.

Figure 1(f) indicates the effect of Se substitution on the layer spacing of 4$H_b$-TaSe$_x$S$_{2-x}$ single crystals. The x-ray diffraction (XRD) patterns of 4$H_b$-TaS$_2$, 4$H_b$-TaSe$_{0.01}$S$_{1.99}$, and 4$H_b$-

TaSe$_{0.7}$S$_{1.3}$ single crystals are shown in Fig. 1(e), in which only (00l) reflections were observed, suggesting the *c* axis is perpendicular to the surface of the crystal. With increasing Se, the diffraction peaks distinctly shift to lower angles, reflecting the crystal expansion induced by Se substitution. That is to say, the Se substitution in 4*H*$_b$-TaS$_2$ will effectively increase the layer spacing. The Atomic Force Microscope (AFM) topography and scanning Kelvin probe force microscopy (SKPM) surface potential images of 4*H*$_b$-TaS$_2$, 4*H*$_b$-TaSe$_{0.01}$S$_{1.99}$, and 4*H*$_b$-TaSe$_{0.7}$S$_{1.3}$ are further given in Fig. S2 and Figs. 1(g)-1(l). For the 4*H*$_b$-TaSe$_x$S$_{2-x}$, a consequence of the T, H, T, stacking pattern is that there are two cleavage planes: T or H surfaces as shown in Figs.1(g), 1(j) and S1(a).

Due to the different electronic properties, the 1T- (1H-) layers can be clearly identified by the relatively high (low) surface potentials in the SKPM images. The surface potential difference between 1T-layer and 1H-layer are ~400 mV of 4*H*$_b$-TaS$_2$ and 4*H*$_b$-TaSe$_{0.01}$S$_{1.99}$ [Figs. 1(i) and S2(c)], which is consistent with previous reports that the work function of the 1H-layer is 5.6 eV while the work function of the 1T-layer is 5.2 eV [36]. With increasing Se, a sizable decrease/increase of the $E_F$ of the 1T/1H-layers can be induced [37], in accordance with the reduced surface potential difference (~370 mV) in 4*H*$_b$-TaSe$_{0.7}$S$_{1.3}$ as shown in Figs. 1(l) and S2(f). The increased interlayer distance and reduced surface potential difference makes 4*H*$_b$-TaSe$_{0.7}$S$_{1.3}$ an ideal platform to investigate interfacial charge transfer of electrons of 1*T*- and 1*H*-TaS$_2$ layers and spatial distributions of correlated electrons in the flat CDW band of 1T-layer.

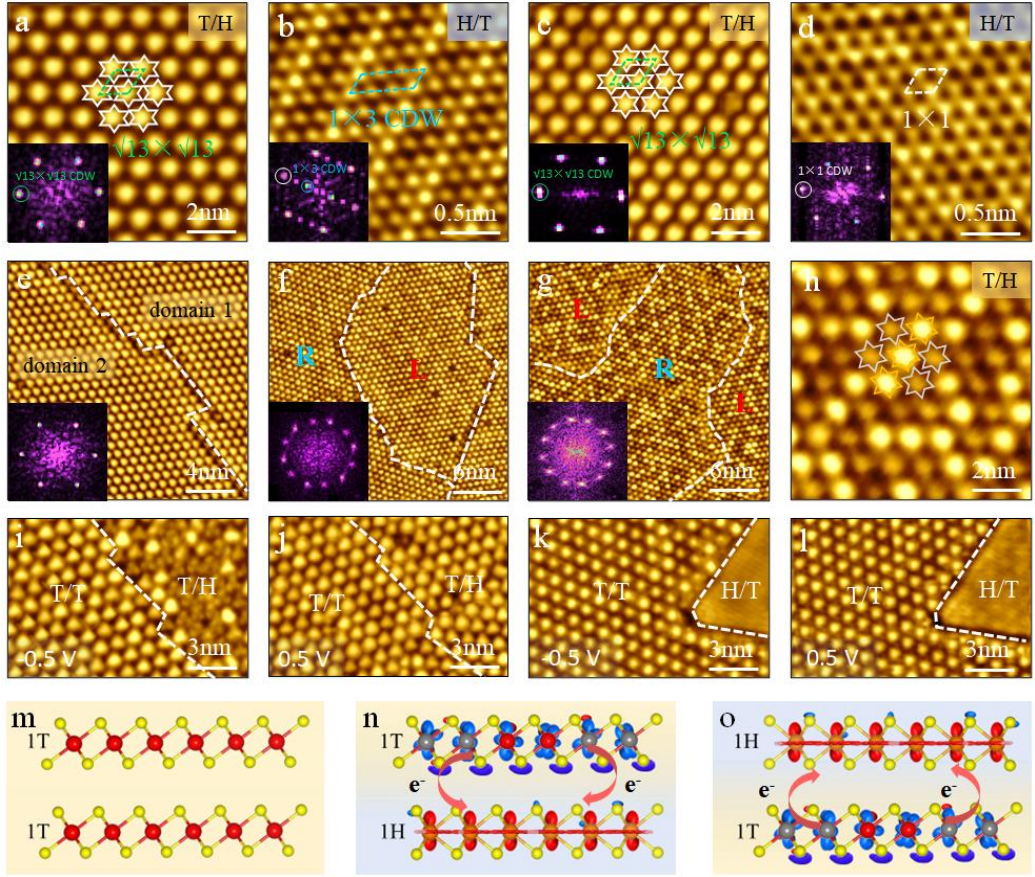

**Figure 2. STM measurements of $4H_b$-TaSe$_{0.01}$S$_{1.99}$ and $4H_b$-TaSe$_{0.7}$S$_{1.3}$.** (a-d) STM images and corresponding FFT patterns of 1T-layer (a,c) and 1H-layer (b,d) of $4H_b$-TaSe$_{0.01}$S$_{1.99}$ and $4H_b$-TaSe$_{0.7}$S$_{1.3}$. The commensurate $\sqrt{13}\times\sqrt{13}$ CDW of 1T-layer are preserved in both (a) and (c), while the 3×3 CDW of 1H-layer are most suppressed to 1×3 CDW($4H_b$-TaSe$_{0.01}$S$_{1.99}$) and negligible CDW ($4H_b$-TaSe$_{0.7}$S$_{1.3}$), respectively. (e) Large-scale STM image and corresponding FFT pattern of the 1T-layer in $4H_b$-TaSe$_{0.01}$S$_{1.99}$. Only very few of CDW phase domain walls are observed. (f,g) Large-scale STM images and corresponding FFT patterns of 1T-layer of $4H_b$-TaSe$_{0.7}$S$_{1.3}$ taken with positive (f) and negative (g) bias. Both CDW phase and chiral domain walls are observed. (h) A zoom-in STM image from (g), showing the distributed bright SoD (orange stars) and dark SoD clusters (gray stars). (i-l) STM images of 1T/1T, 1T/1H and 1H/1T stacking regions in $4H_b$-TaSe$_{0.7}$S$_{1.3}$ taken with positive (i,k) and negative (j,l) bias. The charge transfer at the 1T/1H heterointerface contributed the apparent difference of 1T-layer at occupied and unoccupied states. (m-o) Schematic model of interlayer charge transfer of 1T/1T, 1T/1H, and 1H/1T stacking order. Blue and red represent the electron depletion and accumulation regions. Imaging parameters: (a) $V_b$ = 0.8 V, $I_t$ = 100 pA; (b, d) $V_b$ = 150 mV, $I_t$ = 300 pA; (c) $V_b$ = 1.0 V, $I_t$ = 100 pA; (e, g) $V_b$ = 500 mV, $I_t$ = 100 pA; (f, h, i-l) $V_b$ = - 500 mV, $I_t$ = 100 pA.

A series of detailed STM images and corresponding Fast Fourier transform (FFT) patterns of $4H_b$-TaSe$_{0.01}$S$_{1.99}$ and $4H_b$-TaSe$_{0.7}$S$_{1.3}$ are displayed in Figs. 2(a)-2(d). It can be seen that the CDW of 1T-layer of $4H_b$-TaSe$_{0.01}$S$_{1.99}$ and $4H_b$-TaSe$_{0.7}$S$_{1.3}$ retains the $\sqrt{13} \times \sqrt{13}$

SoDs patterns as that in the intrinsic $4H_b$-TaS$_2$ sample [29] at positive bias [Figs. 2(a) and 2(c)]. However, for the 1H-layer in $4H_b$-TaSe$_x$S$_{2-x}$, the 3 × 3 CDW is gradually suppressed with the increase of Se concentration [38]. For $4H_b$-TaSe$_{0.01}$S$_{1.99}$, the 3×3 CDW of 1H-layer is most suppressed to 1×3 CDW [Figs. 2(b)], while the 3×3 CDW of 1H-layer is negligible of $4H_b$-TaSe$_{0.7}$S$_{1.3}$, displaying the 1 × 1 periodicity [Figs. 2(d) and S3].

How does Se substitution modify the CCDW order of 1T-layer? Figure 2(c)-(g) displays the large-scale STM images and corresponding FFT patterns of 1T-layer of $4H_b$-TaSe$_{0.01}$S$_{1.99}$ and $4H_b$-TaSe$_{0.7}$S$_{1.3}$. Unlike the pristine $4H_b$-TaS$_2$ sample, a very small amount of Se substitution induces a few CCDW phase domain walls of the 1T-layer of $4H_b$-TaSe$_{0.01}$S$_{1.99}$, as shown in Fig.2(e). Increasing the Se concentration from 0.01 to 0.7 creates more domain walls, and the 1T-surface now splits into domains with varied domain size and sharp boundaries, similar to the so-called "mosaic" phase in $1T$-TaS$_2$ induced by a voltage pulse [39]. In addition, the large chiral CDW domains emergent within the mosaic-like CDW states, with the R- and L-chiral domains separated by white chiral domain walls [Figs. 2(f) and 2(g)]. Intriguingly, under negative bias, two distinguished brightness of SoDs, one bright and one dark can be clearly observed of $4H_b$-TaSe$_{0.7}$S$_{1.3}$ [Figs. 2(g) and 2(h)], while for $4H_b$-TaSe$_{0.01}$S$_{1.99}$, a few dispersed bright SoDs emerge [Fig. S4]. This brightness appearance is not topographic, but may derive from different electron fillings of SoDs, similar to the bright-dark distribution of SoDs on the hole Ti-doped $1T$-TaS$_2$ surface [40].

To determine the origin of these bright and dark SoDs of the 1T-layer, different stacking orders, including 1T/1T, 1T/1H and 1H/1T of $4H_b$-TaSe$_{0.7}$S$_{1.3}$ are investigated in Figs. 2(i)-2(l). Notably, the 1T/1H and 1T/1T stacking regions can be clearly distinguished at the negative bias (occupied state). In contrast to the homogeneously distributed SoDs of 1T/1T region, the number of bright SoDs of 1T/1H region is significantly reduced, revealing a pattern of alternating bright and dark SoDs [Fig. 2(i)]. While, all the SoDs of 1T/1T and 1T/1H exhibit the same shape and brightness at the positive bias (unoccupied state) [Fig. 2(j)]. These observations illustrate that the charge transfer at the 1T/1H heterointerface contributed the apparent difference of 1T-layer at occupied and unoccupied states. This can be further demonstrated by STM images of 1T/1T and 1H/1T stacking regions in $4H_b$-TaSe$_{0.7}$S$_{1.3}$ taken with positive and negative bias in Figs. 2(k) and 2(l). Figures 2(m)-2(o) gives the schematic model of interlayer charge transfer of 1T/1T, 1T/1H, and 1H/1T stacking order, in which the 1H-layer provides a charge reservoir, taking away some electrons from the 1T-layer. However, compared to the 1T/1H overall electron distribution of intrinsic $4H_b$-TaS$_2$, the 1T/1H of $4H_b$-TaSe$_{0.7}$S$_{1.3}$ demonstrates a spatial localized electron-filled configuration with a bright and dark SoDs distribution. We attribute this difference to the Se substitution, which effectively changes the interlayer charge transfer mode of electrons from overall to partial charge transfer [Fig. S5].

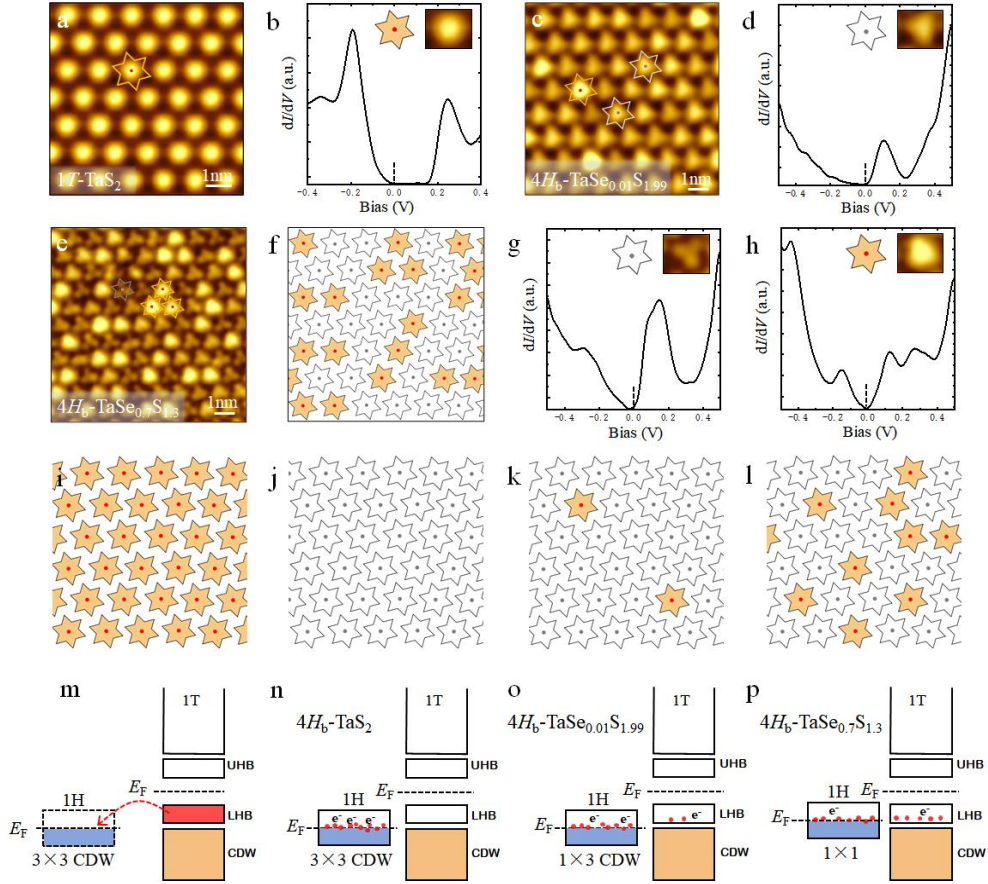

**Figure 3. Electronic states of the 1T-layer of $4H_b$-TaSe$_{0.01}$S$_{1.99}$ and $4H_b$-TaSe$_{0.7}$S$_{1.3}$.** (a,b) Typical STM image (a) and d$I$/d$V$ spectra (b) taken on the pristine $1T$-TaS$_2$. (c,d) STM image (c) and d$I$/d$V$ spectra (d) of the 1T-layer of $4H_b$-TaSe$_{0.01}$S$_{1.99}$. (e,f) Typical STM image (e) and schematic model (f) of the distributed SoDs in the 1T-layer of $4H_b$-TaSe$_{0.7}$S$_{1.3}$. (g,h) Typical d$I$/d$V$ spectra of dark (g) and bright (h) SoDs in (e,f). (i-p) Schematic of distributed bright/dark SoDs (i-l) and band diagrams (m-p) for the interfacial electron transfer in $1T$-TaS$_2$, $4H_b$-TaS$_2$, $4H_b$-TaSe$_{0.01}$S$_{1.99}$ and $4H_b$-TaSe$_{0.7}$S$_{1.3}$. The electron-filled and -void SoDs are marked by the stars with the central red and grey dots, respectively. Different from the intrinsic Mott spectra in $1T$-TaS$_2$, the d$I$/d$V$ spectra of electron-filled and -void SoDs show prominent spectra weight above $E_F$ and symmetric Mott-like spectra weight with small U, respectively. Imaging parameters: (a) $V_b$ = -1V, $I_t$ = 100 pA; (c,e) $V_b$ = -500 mV, $I_t$ = 100 pA.

Next, we turn to site-specific STS spectra measurements on the 1T-layer of $1T$-TaS$_2$, $4H_b$-TaS$_2$, $4H_b$-TaSe$_{0.01}$S$_{1.99}$ and $4H_b$-TaSe$_{0.7}$S$_{1.3}$ to investigate the origin of the bright and dark SoDs, as shown in Fig. 3(a)-3(h). Compared with the electron-hole symmetric gap of the SoDs in pristine $1T$-TaS$_2$ [Fig. 3(b)], the d$I$/d$V$ spectra of the bright triangular SoDs (orange stars) and the dark triangular SoDs (grey stars) on 1T-layer of $4H_b$-TaSe$_{0.01}$S$_{1.9}$ reveal significantly differences. The bright triangular SoDs exhibit symmetric Mott-like V-shaped gap [Fig. S6], while the dark triangular SoDs display an electron-hole asymmetric single pronounced peak above the $E_F$ [Fig. 3(d)], in almost coincidence with the d$I$/d$V$ spectra of the 1T-layer of typical 4$H_b$-TaS$_2$ [29]. The coexistence of these two distinct spectra of bright and dark triangular SoDs of 4$H_b$-TaSe$_{0.01}$S$_{1.9}$ has not been reported in 4$H_b$-TaS$_2$. The emergence of bright SoDs possibly be related to a small electron filling of flat-band of 1T-layer caused by the reduction of interlayer charge transfer, and the V-shaped spectral characteristic is due to the screening of the Coulomb interaction by the below metal 1H-layer.

Typical STM images and schematic diagrams of the 1T-layer of 4$H_b$-TaSe$_{0.7}$S$_{1.3}$ are given in Figs. 3(e) and 3(f), illustrating an alternating pattern of bright triangular and dark three-petal-flower SoDs. These variations reflect the different electron-fillings of the SoDs of the 1T-layer, with the brighter triangular SoDs (orange stars) and three-petal-flower SoDs (white stars) indicating the central electron-filled and -void SoDs, respectively. The three-petal-flower SoDs demonstrate that one Mott electron per SoD is transferred to the 1H-layer. The corresponding d$I$/d$V$ spectra of the dark three-petal-flower and the bright triangle SoDs of 4$H_b$-TaSe$_{0.7}$S$_{1.3}$ are shown in Figs. 3(g) and 3(h), respectively. The dark three-petal-flower SoDs featured by spectral weight transfer prominent spectra weight above $E_F$ with suppression of the LHB, similar to the d$I$/d$V$ spectra of dark triangular SoDs of 4$H_b$-TaSe$_{0.01}$S$_{1.99}$. The apparent difference between the dark triangle SoDs (brighter center) of 4$H_b$-TaSe$_{0.01}$S$_{1.99}$ and the dark three-petal-flower SoDs (darker center) of 4$H_b$-TaSe$_{0.7}$S$_{1.3}$ may originate from the different electron-filling factors of the Mott electrons and/or the influence of the 1×3 CDW /1×1 of metallic 1H-layer. The brighter triangles SoDs show a symmetric Mott-like spectra and a finite the density of states (DOS) at $E_F$ with small U. These diverse electronic states emerging from different SoDs configurations indicate the band filling as a pivotal factor in the complexity of 4$H_b$-TaSe$_{0.01}$S$_{1.99}$ and 4$H_b$-TaSe$_{0.7}$S$_{1.3}$.

Schematic of distributed bright/dark SoDs and band diagrams for the interfacial electron transfer in 1$T$-TaS$_2$, 4$H_b$-TaS$_2$, 4$H_b$-TaSe$_{0.01}$S$_{1.99}$ and 4$H_b$-TaSe$_{0.7}$S$_{1.3}$ are illustrated in Figs. 3(i)-3(p). For the 1$T$-TaS$_2$, each SoD contains one unpaired electron (Mott) of the central Ta atom in a half-filled metallic band. The half-filled band splits further UHB and LHB and opening a Mott gap due to the large onsite Coulomb interaction U, corresponding to the overall n = 1 state [Figs. 3(i) and 3(m)]. For the intrinsic 4$H_b$-TaS$_2$, the work function difference between 1T- and 1H-layers results in a substantial interface charge transfer, leading to the heavily hole-doped 1T-layer, where 1e per SoD is transferred from the 1T to the 1H layer. Interface charge transfer contributes to the overall empty state (n = 0) and depleted flat band above the $E_F$ of the 1T-layer in $4H_b$-TaS$_2$ [Figs. 3(j) and 3(n)]. In line with this, the $1T$-TaS$_2$ layer in $4H_b$-TaS$_2$ shows a hole-doped Mott state with the spectral peak formed just above $E_F$.

With the substitution of Se, the interlayer distance increases, inducing a decrease in interlayer hybridization and interlayer charge transfer [Fig. S5], where the flat band of the 1T-layer changes from completely empty to partially occupy. For the $4H_b$-TaSe$_{0.01}$S$_{1.99}$, the Se substitution causes the emergence of very little electrons in the flat band, which appear as bright SoDs (central electron-filled) in real-space [Figs. 3(k) and 3(o)], and the rest of the dark SoDs are essentially coincident with the intrinsic $4H_b$-TaS$_2$. When the Se substitution concentration is 0.7, the charge transfer decreases from 1e to 0.7e [35] and the flat band filling factor increases accordingly from 0 to 0.3e [Figs. 3(l) and 3(p)], which is consistent with the results of the partially filled flat bands of ARPES [41]. Once the flat band starts to host a portion of the electron, the peak of the flat band shifts towards $E_F$, and a considerably small Coulomb interaction U would be sufficient to cause a band splitting. The partially filled flat bands contribute a spatial fractional electron-filling configurations with the distributed electron-filled and -void SoD clusters of 1T-layer.

In addition, another factor affecting the size of the charge transfer and the real-space electron distribution of the T-layer may be the presence or absence of 3 × 3 CDW of the H-layer. We suggest that the 1 × 1 CDW has a higher work function than the 3 × 3 CDW of the 1H-layer, so fewer electrons are transfer from the 1T-layer, and that the 1 × 1 CDW is more metallic, providing stronger Coulomb screening of the 1T-layer. These observations imply that interlayer charge transfer can be modulated by the Se substitution and CDW substrate of 1H-layer, and the band filling is crucial to determining the spatial fractional electron-filling configurations of 1T-layer.

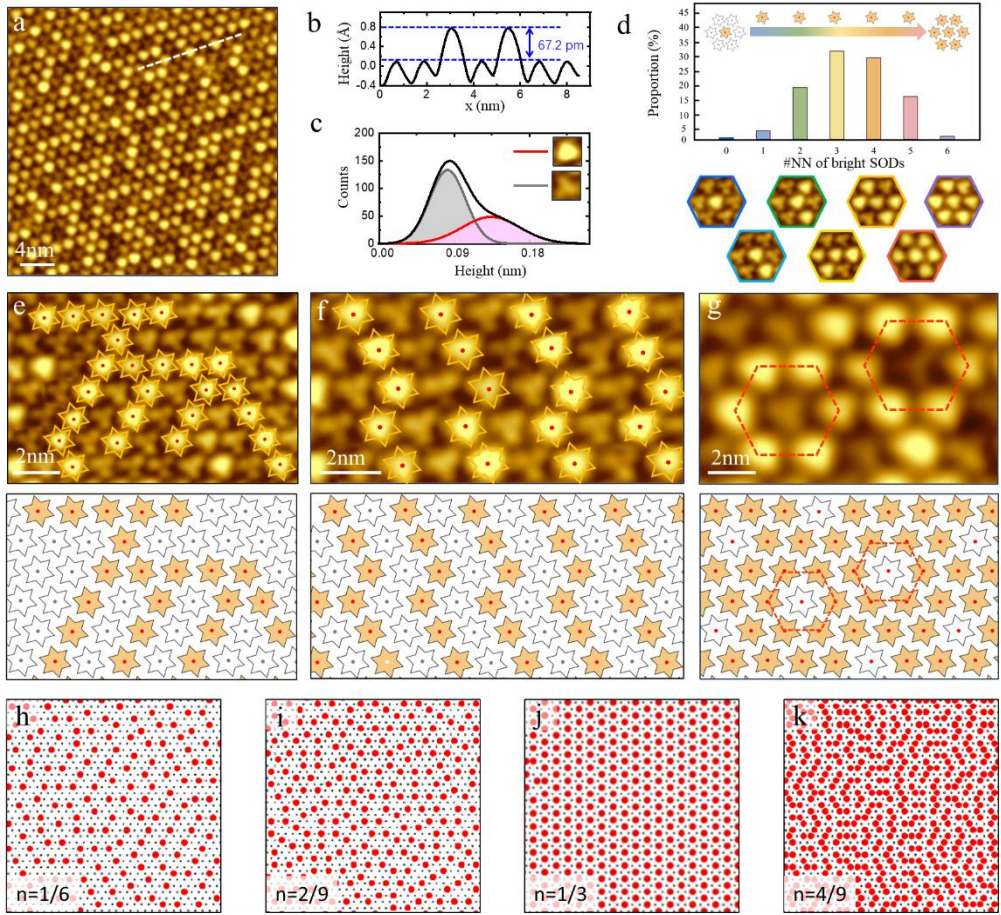

**Figure 4. Correlated electron distributions of the 1T-layer of $4H_b$-TaSe$_{0.7}$S$_{1.3}$.** (a) Representative large-scale STM image of electron distributions. (b) Apparent height distribution curves of (a). The bright SoDs account for ~30% of all SoDs in the 1T-layer of $4H_b$-TaSe$_{0.7}$S$_{1.3}$. (c) The height profile along the dashed line in (a) shows a height difference of ~67.5pm between the bright and dark SoDs. (d) Histogram counting nearest-neighbor SoDs having different brightness, averaged over fifteen STM images (~7000 SoDs). The illustrations show the configuration types associated with the histogram values, taking the central SoDs of the seven-SoDs plaquette as a reference. All other configurations can be derived by arrangement permutations. (e-g) High-resolution STM images (top) and their schematic models (bottom) of the observed strip-like (e), short-range $1 \times 2$ SoDs (f) and anti-seven magic number (g) electron configurations. (h-k) Reproduced electrons patterns (marked by red dots) in the triangular lattice at fractional fillings of n=1/6 (h), n=2/9 (i), n= 1/3 (j) and n= 4/9 (k) simulated based on a Coulomb gas model [42]. Imaging parameters: (a, e-g) $V_b$ = - 500 mV, $I_t$ = 100 pA.

The spatial distribution of the bright (electron-filled) SoDs of the 1T-layer in $4H_b$-TaSe$_{0.7}$S$_{1.3}$ was further investigated by a large number of STM images, as shown in Fig. 4a and Fig. S8. It can be seen that the apparent height of bright (electron-filled) and dark (electron-void) SoDs is different in the occupied state [Fig. 4(b)]. Based on these differences, we can estimate the overall electron filling factor in the 1T-layer to be n = 1/3, which reflects the interlayer electron transfer. Unlike the overall uniform electron transfer in $4H_b$-TaS$_2$, $4H_b$-TaSe$_{0.7}$S$_{1.3}$ demonstrates spatially localized electron transfer. Careful observation reveal that the spatial distributions of the bright SoDs are not random, but exhibits the correlated order, such as stripe-like, short-range 1×2 SoDs and anti-seven magic number electron configurations [Figs. 4(e)-4(f)], which reflects the complex interactions between the electrons in the 1T-layer. In order to analyze the nearest-neighbor interactions among the bright SoDs, the assembled form of the nearest-neighbor of bright SoDs is counted. Histogram counting nearest-neighbor SoDs indicate that the bright SoDs tend to form a 4/5-SoDs configuration rather than preferring to exist in isolation or to form a 7-SoDs configuration [Fig. 4(d)]. The stripe-like SoDs and SoDs clusters configurations imply the existence of nearest-neighbor attractive interactions among bright SoDs, while the short-range 1×2 SoDs and anti-seven magic number (7-SoDs) electron configurations suggest that there are also conformation-dependent repulsive interactions.

Previous studies using STM and scanning microwave impedance microscopy (SMIM) techniques confirmed the existence of generalized Wigner crystal state with different fractional fillings in twisted system, which can be described by a simple Coulomb gas model [43-45]. The ordered patterns of Monte Carlo (MC) simulations based on the Coulomb gas model with different fractional fillings are shown in Figs. 4(h)-4(k). Compared to the ordered patterns of MC simulations, our electron distributions demonstrate some different patterns at the same fractional filling. This difference derives from the fact that only the long-range Coulomb repulsion is considered in the Coulomb gas model, while both short-range attractive and configuration-dependent repulsive interactions are present in the electron distribution of $4H_b$-TaSe$_{0.7}$S$_{1.3}$. Our electron-filling configurations cannot be described by a simple Coulomb gas model, but by an extended Hubbard model, which takes into account near-neighbor interactions, including attractive and repulsive forces, in addition to the U [46,47]. Our work firstly investigates the real-space fractional electron-filling configurations, suggesting the existence of complex attractive and repulsive interactions between electrons. The origin of the theses complicated interactions is still unclear and needs to be further studied.

**Conclusion**

$4H_b$-TaS$_2$ interleaves the Mott-insulating state of $1T$-TaS$_2$ and the putative spin liquid it hosts together with the metallic state of $2H$-TaS$_2$ and the low temperature superconducting phase it harbors, providing a good platform for the investigation of the competition/ cooperation between superconductivity and charge-density-wave (CDW) order. Recent studies illustrated the differences difference in spectroscopic between the Kondo resonated 1T/1H bilayer and the hole-doped Mott insulator $4H_b$-TaS$_2$, implying that doping levels in the Mott insulating 1T-layer can be effectively modulated through the interlayer coupling. With regard to the tuning means for the interlayer coupling, the isovalent Se substitution acts as an effective method，which does not introduce extra electron or hole into the system.

We investigated in detail the effects of the Se substitution on the work-function, CDW order, the interfacial electrons transfer between $1T$- and $1H$-TaS$_2$ layers, and the electron-filled of the flat band of the 1T-layer of superlattice $4H_b$-TaSe$_x$S$_{2-x}$. Significantly, compared with pristine $4H_b$-TaS$_2$, Se substitution increases the interlayer distance between the 1T- and 1H-layers, accompanied by a decrease in charge transfer and a decrease in hybridization. The decrease of charge transfer contributes a spatial fractional electron-filling configurations with the distributed electron-filled and -void SoD clusters of 1T-layer. STS spectra results further exhibit respective prominent spectra weight above $E_F$ and symmetric Mott-like spectra of electron-filled and -void SoD clusters. Our work provides insights into understanding correlated electrons of flat band of the superlattice $4H_b$-TaSe$_x$S$_{2-x}$, but the effect on chiral superconductivity is not yet known. The influence of different substitution ratio on CDW and superconductivity can be explored at lower temperatures in the future to realize more exotic quantum states. Furthermore, the complex interactions between localized electrons of the 1T-layer imply the potential for novel electronic states, providing a simple platform for manipulating these interactions to control the electron-correlation-related quantum states.

**Materials and methods**

The high-quality $4H_b$-TaSe$_x$S$_{2-x}$ single-crystal samples were grown by the chemical vapor transport (CVT) method with iodine as the transport agent. Stoichiometric amounts of high-purity elements Ta (99.9%), S (99.9%), and Se (99.95%) were mixed. The doping has been uniformly and precisely controlled in high-quality samples. The $4H_b$-TaSe$_x$S$_{2-x}$ crystals were cleaved at room temperature in ultrahigh vacuum at a base pressure of $2\times10^{-10}$ Torr, and directly transferred to the cryogen-free low-temperature STM system (PanScan Freedom-LT, RHK). Chemically etched Pt-Ir tips were used and calibrated on a clean Ag(111) for STM and STS measurements. All the STM/STS measurements were performed at the base temperature of ~9 K. The differential conductance (d$I$/d$V$) spectra were measured in constant-height mode using standard lock-in techniques. Gwyddion was used for STM data analysis.

## Acknowledgments

This project was supported by the National Key R&D Program of China (MOST) (Grant No. 2023YFA1406500), the National Natural Science Foundation of China (NSFC) (No. 21622304, 61674045, 11604063, 11974422, 12104504), the Strategic Priority Research Program (Chinese Academy of Sciences, CAS) (No. No. XDB30000000), and the Fundamental Research Funds for the Central Universities and the Research Funds of Renmin University of China [No. 21XNLG27 (Z.C.), No. 22XNH095 (H.D.)]. Y.Y. Geng was supported by the Outstanding Innovative Talents Cultivation Funded Programs 2023 of Renmin University of China.

## Author contributions

Y. G., J. G. and Z. C. performed the STM experiments and analysis of STM data. S. Mi., M. W., H. D., R. Xu., F. Pang., K. Liu., S. Wang., helped in the experiments. F. M., and H. L. provided the sample. W. J and Z. C. wrote the manuscript with inputs from all authors.

## Competing Interests

The authors declare no competing financial interests.

## Data Availability

The authors declare that the data supporting the findings of this study are available within the article and its Supplementary Information.